\documentclass[10pt, conference]{IEEEtran}
\usepackage{pstricks}
\usepackage{adjustbox}
\usepackage{listings}
\usepackage{float}
\usepackage{svg}
\usepackage{amssymb}
\usepackage{amsmath}
\usepackage{subfig}
\usepackage{graphicx}
\usepackage{url}
\usepackage{tabularx}
\usepackage[htt]{hyphenat}
\usepackage{algorithm}
\usepackage[noend]{algpseudocode}
\usepackage{tikz}
\usetikzlibrary{shapes,arrows,positioning}
\usepackage{epstopdf}
\usepackage{pgfplots}
\usepackage{tkz-graph}
\usepackage{amssymb}
\usepackage{pifont}
\usepackage[flushleft]{threeparttable} % http://ctan.org/pkg/threeparttable
\usepackage[bookmarks=false,hidelinks]{hyperref}
\usepackage{url}
\usepackage{footnote}
\makesavenoteenv{tabular}
\makesavenoteenv{table}

\usepackage{array}
\newcolumntype{L}[1]{>{\raggedright\let\newline\\\arraybackslash\hspace{0pt}}m{#1}}
\newcolumntype{C}[1]{>{\centering\let\newline\\\arraybackslash\hspace{0pt}}m{#1}}
\newcolumntype{R}[1]{>{\raggedleft\let\newline\\\arraybackslash\hspace{0pt}}m{#1}}

\newcommand{\thething}{liOS}

\def\BibTeX{{\rm B\kern-.05em{\sc i\kern-.025em b}\kern-.08emT\kern-.1667em\lower.7ex\hbox{E}\kern-.125emX}}

\makeatletter
\def\BState{\State\hskip-\ALG@thistlm}
\makeatother

\algrenewcommand\algorithmicforall{\textbf{foreach}}
\algrenewcommand\algorithmicindent{.8em}

\newcommand{\textapprox}{\raisebox{-.75ex}{\textasciitilde}}

\begin{document}
\title{\thething{}: Lifting iOS Apps for Fun and Profit}

\author{
\IEEEauthorblockN{Julian Sch\"{u}tte and Dennis Titze}
\IEEEauthorblockA{Fraunhofer AISEC\\
Garching, Germany\\
Email: \{julian.schuette,dennis.titze\}@aisec.fraunhofer.de}
}

%
%\authorrunning{J. Sch\"{u}tte and D. Titze}
%\authorrunning{BLINDED FOR REVIEW}
%\institute{Fraunhofer AISEC
%\email{julian.schuette@aisec.fraunhofer.de}\\
%\url{http://www.aisec.fraunhofer.de}}

%\institute{BLINDED FOR REVIEW
%\email{}\\
%\url{}}

\maketitle

\begin{abstract}
Although iOS is the second most popular mobile operating system and is often considered the more secure one, approaches to automatically analyze iOS applications are scarce and generic app analysis frameworks do not exist. This is on the one hand due to the closed ecosystem putting obstacles in the way of reverse engineers and on the other hand due to the complexity of reverse engineering and analyzing app binaries. Reliably lifting accurate call graphs, control flows, and data dependence graphs from binary code, as well as reconstructing object-oriented high-level concepts is a non-trivial task and the choice of the lifted target representation determines the analysis capabilities. None of the various existing intermediate representations is a perfect fit for all types of analysis, while the detection of vulnerabilities requires techniques ranging from simple pattern matching to complex inter-procedural data flow analyses. We address this gap by introducing \thething{}, a binary lifting and analysis framework for iOS applications that extracts lifted information from several frontends and unifies them in a ''supergraph'' representation that tolerates missing parts and is further extended and interlinked by \thething{} ''passes''. A static analysis of the binary is then realized in the form of graph traversal queries, which can be considered as an advancement of classic program query languages. We illustrate this approach by means of a typical JavaScript/Objective-C bridge, which can lead to remote code execution in iOS applications.
\end{abstract}

% considered as the vs. considered the https://english.stackexchange.com/questions/154347/difference-between-considered-to-be-and-considered-as

% -- begin of section to replace with generated code
%  \begin{CCSXML}
% <concept_id>10002978.10003022.10003465</concept_id>
% <concept_desc>Security and privacy~Software reverse engineering</concept_desc>
% <concept_id>10002978.10003022.10003028</concept_id>
% <concept_desc>Security and privacy~Domain-specific security and privacy architectures</concept_desc>
% <concept_significance>300</concept_significance>
%<concept>
%<concept_id>10011007.10011006.10011050.10011017</concept_id>
%<concept_desc>Software and its engineering~Domain specific languages</concept_desc>
%<concept_significance>100</concept_significance>
%</concept>

%\ccsdesc[500]{Security and privacy~Software reverse engineering}
%\ccsdesc[300]{Security and privacy~Domain-specific security and privacy architectures}
%\ccsdesc[100]{Software and its engineering~Domain specific languages}
% -- end of section to replace with generated code

% replace with your keywords
%\keywords{mobile security; binary lifting; code property graph; reverse engineering; static analysis; iOS} 

%\maketitle
%\tableofcontents

\section{Introduction}

iOS is the second most spread mobile operating system and has ever since enjoyed a reputation as the ''more secure'' system, compared to Android. While this might actually be true when considering the security features of the iOS hardware and software stack, a substantial part of this reputation originates from the fact that the app ecosystem is largely under exclusive control of Apple. The bar for distributing malware is much higher due to the obligation to enroll in the Apple developer program, the AppStore review process, and Apple's ability to centrally revoke an app. However, malware is not the only threat to users' security and privacy, and for iOS it might even be the less relevant one.

The closed nature of the app ecosystem and the fact that iOS applications are much harder to reverse engineer -- let alone repackage -- than Android apps has drawn the attention of many researchers and hackers away from iOS to platforms that are easier to assess and attack. But iOS applications are not necessarily more secure than their Android equivalents. In contrast to Android apps which mostly consist of memory-safe bytecode and are assigned a separate user ID per app, iOS apps typically run under the same user account \texttt{mobile} and are confined by the Sandbox which enforces mandatory access control profiles on them -- a mechanism which has been shown to be flawed in the past \cite{Deshotels2016}. For users, the internals of an iOS app remain highly unclear and trust in an app relies exclusively on the AppStore review process whose details are not publicly known and which has been circumvented in the past \cite{wang2013jekyll}. Thus, ways to automatically analyze iOS apps for vulnerabilities are urgently needed to increase transparency for the user and to build trust in the ecosystem.

Precise automated static analysis of binary iOS applications is however not trivial and only few contributions have been made so far by the research community \cite{EgeleM.KruegelC.KirdaE.Vigna2011,Feichtner2018}. First, the majority of iOS apps are not easily accessible due to the FairPlay DRM encryption which is supposed to prevent apps from being copied. Although not insurmountable, the mechanism is a considerable hurdle for researchers to get their hands on a large corpus of iOS apps, as discussed by Orikogbo et al. \cite{crios2016}. Further, iOS apps do not come in form of easily analyzable bytecode, but in form of native ARM binaries that  make it more difficult and error-prone to recover high-level information such as function calls and variable aliases. Above-cited research has shown that it is still possible to detect specific vulnerabilities by analyzing the app binary but up to date, there is no generic static analysis framework available that allows to detect arbitrary vulnerability pattern in iOS apps. Specifically, the following challenges have not been satisfactorily solved:
\begin{itemize}
  \item Lifting iOS apps to a representation that covers both the semantics of low-level assembly as well as high-level object oriented concepts
  \item A thorough approach to take into account the semantics of the Objective-C/Swift runtime
  \item A generic way to statically analyze iOS apps for different types of vulnerabilities
\end{itemize}

In this paper, we address these challenges and propose \thething{}, a framework for static binary analysis that can detect configuration-, control flow-, and data flow-based vulnerabilities. \thething{} is extensible in that it accepts input from various \emph{frontends} and combines them in a single graph-based representation, called the \emph{supergraph}. While some frontends operate directly on the disassembly, others lift the aarch64 binary to the LLVM intermediate representation (IR) to support further analysis techniques from the existing LLVM ecosystem, or extract higher-level information such as class hierarchy, methods, and variables from the binary. The graph-based representation allows us to combine the output of these frontends into one unified representation, linking information from all frontends with each other. This is especially useful when dealing with an incomplete representation -- a recurring problem with binary lifting, which is never fully accurate for applications of realistic size and complexity. 
Further, it allows us to map the implementation of static analysis methods in \thething{} to the problem of finding a \emph{graph traversal}, i.e. a set of paths in the supergraph with specific properties. This approach decouples the actual artifact (in our case, an iOS app) from the overall analysis framework and allows to plug in further analysis techniques at a later time by simply adding graph traversals. 
%As previous publications only scarcely cover the details of lifting iOS apps into a representation that is suited for static analysis, we will guide the reader through the process of reverse engineering an iOS binary and shed light on specific design choices to make. 
We give details on the supergraph construction and illustrate how typical static analysis problems can be solved with graph traversals. Through a typical remote code execution vulnerability, we illustrate how \thething{} is able to discover complex vulnerabilities in iOS binaries.

% The next section provides background information on the structure and peculiarities of iOS application, the contained Mach-O binaries and the Objective-C runtime against which they are compiled and linked. \autoref{sec:ir} discusses different options of intermediate representations (IR). 
\autoref{sec:reveng} guides through the overall process of automated reverse engineering and analysis of iOS apps, conducted by \thething{}. \autoref{sec:supergraph} introduces the concept of the supergraph and \autoref{sec:framework} explains how \thething{} uses it to analyze iOS apps. We show how \thething{} detects even complex vulnerabilities in iOS apps and discuss its practical application in \autoref{sec:evaluation}. \autoref{sec:relawo} discusses related work and \autoref{sec:conclusion} concludes the paper.

\section{The Reverse Engineering Process}
\label{sec:reveng}

\begin{figure*}
  \centering
  \includegraphics[width=.75\textwidth]{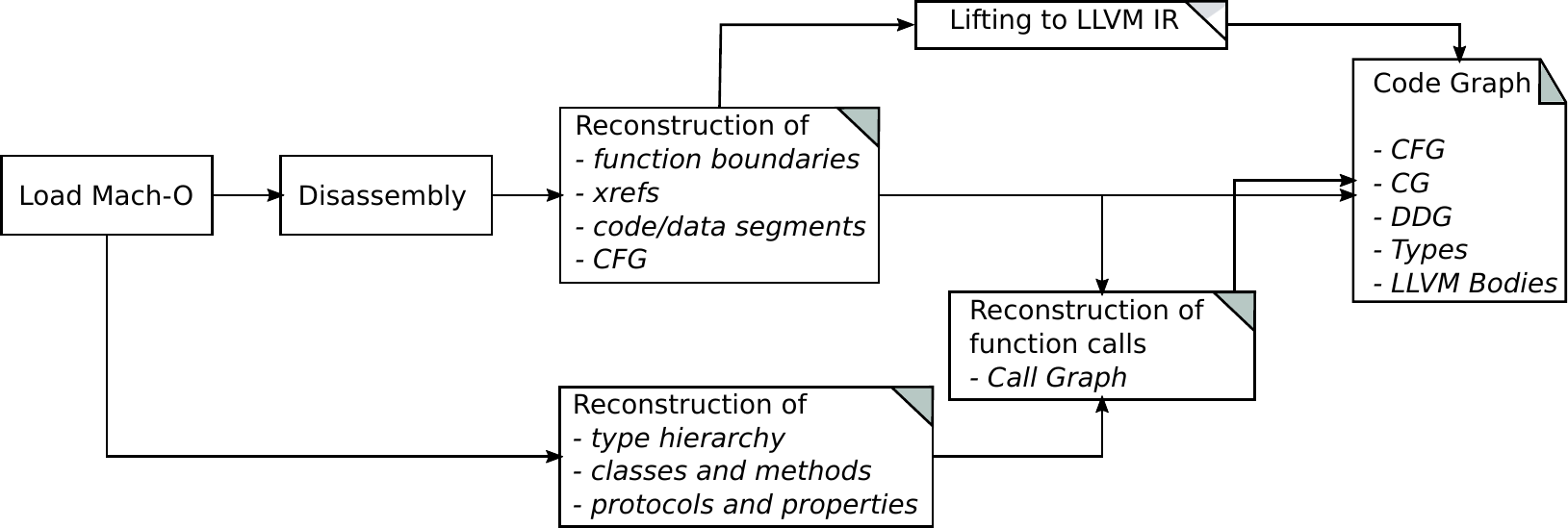}
  \caption{Lifting and reverse engineering process}
  %\Description[Lifting process]{Lifting an reverse engineering process}
  \label{fig:process}
\end{figure*}
Automated reverse engineering of iOS apps is a complex process involving several steps. \autoref{fig:process} depicts this process and shows the contributions of \thething{} (marked corners). In this section, we will walk through this process and explain how \thething{} converts a binary iOS application into a representation suited for program analysis.

\subsection{Unzipping the IPA and Loading the Binary}

We start with a decrypted \texttt{.ipa} file that is either directly exported as an archive from XCode or has been dumped from a physical device.
When installed from the Apple AppStore, Mach-O sections will be encrypted with the public key that is assigned to the Apple account associated with the physical device, whose corresponding private key is managed by the Secure Enclave TEE on the phone. This mechanism is part of Apple's FairPlay DRM and ways to circumvent it are known since 2008. As decryption of sections is already done by the binary loader when mapping sections to memory segments, it is easy enough to load the app into memory, dump its segments in clear text and re-assemble an unencrypted binary. Orikogbo et al. have shown in \cite{crios2016} that this process can be fully automated and although their success rate when decrypting the binary is only about 51\%, our experience is that a much higher ($>$ 95\%) success rate can be achieved by using more recent devices and the Frida DBI framework\footnote{\nolinkurl{https://github.com/AloneMonkey/frida-ios-dump}}.

The actual analysis process of \thething{} begins with extracting the Mach-O binary from the zip-compressed .ipa file, along with further files which are relevant for the analysis at a later time, such as \texttt{Info.plist}. The aarch64 binary is extracted from the fat Mach-O and passed to the loader of \thething{} which parses the binary's load commands to reconstruct segments (\texttt{LC\_SEGMENT}), symbol tables (\texttt{LC\_SYMTAB} and \texttt{LC\_DYSYMTAB}), and function boundaries (\texttt{LC\_FUNCTION\_STARTS}).
% Also: LC_DATA_IN_CODE and LC_MAIN (entrypoint)
The detection of function boundaries in a binary blob in the absence of debug symbols is a notoriously difficult problem and modern disassemblers apply heuristic approaches such as \textsc{Byteweight} \cite{Bao2014}, FID \cite{Wang2017} or FLIRT \cite{Flirt} to determine where a function starts. Luckily, in contrast to PE and ELF binaries, the Mach-O format includes an \texttt{LC\_FUNCTION\_STARTS} load command in its header which points to the address of a list of function start pointers. This list is implemented as a zero-terminated sequence of ULEB128-encoded \cite[pp. 221]{dwarf2017} addresses indicating the start of all functions, from the lowest to the highest address. While this function list is mainly used for producing meaningful output for debuggers or stack traces, it is not to be confused with debug symbols which would get removed from the binary by the \texttt{strip} command. The function list remains intact even in stripped binaries and allows us to precisely and efficiently reconstruct function boundaries from any iOS application.

%\todo{Dennis?: describe reconstruction of Symbol table, dynamic symbol table and string table}

% \subsection{Reconstruction of global and stack variables}

% \todo{Julian}

\subsection{Class hierarchies and Selectors}

%\todo{Swift dispatch: https://developer.apple.com/swift/blog/?id=27}

iOS apps are written in either Swift of Objective-C which are both executed in the Objective-C binary runtime, but may also include C/C++ libraries. A subset of the functions in the aforementioned function list will thus map to methods of Objective-C or Swift classes and reconstructing this mapping along with a correct class hierarchy is essential for creating a clean call graph. As the Objective-C runtime needs precise information about the class hierarchy to properly resolve method calls, this information must always be contained in the Mach-O file and we can extract it from the file's sections.

\begin{figure}[tb]
  \centering
  \includegraphics[width=.7\linewidth]{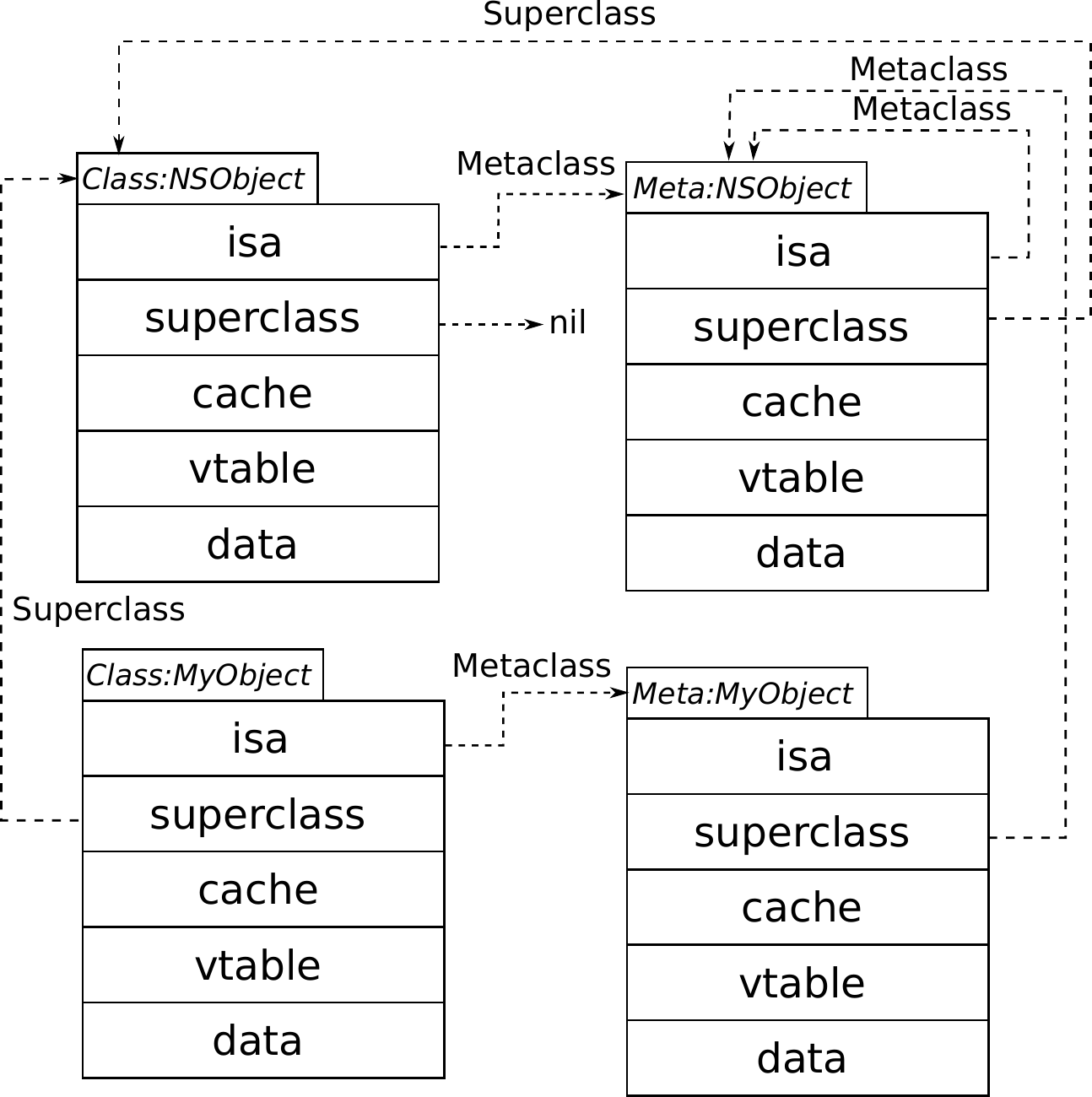}
  \caption{A class- and meta-class hierarchy\label{fig:classes}}
\end{figure}

Section \texttt{\_\_objc\_classlist} contains a list of pointers to \texttt{class\_t} structs, describing the classes contained in the program by their superclass, meta class, size, protocols, methods, instance variables, and properties. Section \texttt{\_\_objc\_classref}, in contrast, contains a list of \texttt{classref\_t} structs describing all classes \emph{used} by the program at runtime. %This list is however populated by the loader at runtime and is not available for a static analysis.
In the Objective-C runtime (i.e. in Swift and Objective-C likewise), every struct with an \texttt{isa} pointer to a \texttt{class\_t} struct is considered a class. The \texttt{isa} pointer indicates the \emph{meta-class} of the class, i.e. the object providing the methods and properties operating on the class itself -- similar to the \texttt{Class} object in Java. Just as every class can have a superclass, every meta-class has a superclass, too. The top of the concrete class hierarchy is indicated by a \texttt{nil} \texttt{superclass} pointer (typically in \texttt{NSObject}), while the top of the meta-class hierarchy is indicated by a cycle, i.e. an \texttt{isa} pointer pointing to the meta-class itself. It is one of Objective-C's quirks that the topmost meta-class has a \texttt{superclass} pointer to its corresponding concrete class and that accessing the \texttt{class} method of a \texttt{Class} object does not provide its meta-class, but rather the \texttt{Class} object itself. 
\autoref{fig:classes} illustrates this possibly at first confusing constellation
%case of the two connected class hierarchy trees of classes and meta-classes.% ie. [NSString class] does not return the meta-class of NSString, but again NSString. Rather use object_getClass

Besides classes, the Objective-C runtime supports \emph{protocols}, which are imple\-mentation-less and class-independent definitions of methods and properties. Me\-thods declared by a protocol are marked required or optional, and classes adopting the protocol must provide implementations for the former, but may omit any optional method. Protocols support inheritance and classes may adopt any number of protocols. However, in contrast to classes, there is no such thing as meta-protocols -- in fact, the \texttt{Protocol} object type extends the base class \texttt{NSObject}. In many cases, the iOS system APIs specify protocols to be implemented by application classes, e.g. to receive data from an API call in a callback function. Thus, knowing the protocols implemented by an app helps reverse engineers understanding the used APIs and functionality of the app. \thething{} extracts protocol information from the respective Mach-O sections and includes it in the type hierarchy. 

% SELECTORS (i.e. Methods)
Methods of Objective-C classes are identified by \emph{selectors}. Their names are listed as null-terminated strings in section \texttt{\_\_objc\_methname}, referenced by a list of pointers in section \texttt{\_\_objc\_selref}. As methods are referenced by selectors, rather than direct pointers, the implementation of a method is only loosely coupled to its name and its class, which allows the Objective-C runtime to dynamically manipulate methods (``method swizzling'') and load code. This is great for developers, but a challenge for secure programming and static analysis. As method calls in Objective-C are not implemented as direct \texttt{BL/BLX} branches to specific addresses but rather as a \emph{message} containing the desired object (``IMP'') and method (``SEL'') that is sent to the Objective-C runtime via the \texttt{obj\_msgSend} function, a naively constructed call graph is almost exclusively centered around that function and does not correctly reflect method calls.
% See https://www.cocoawithlove.com/2010/01/what-is-meta-class-in-objective-c.html
% https://stackoverflow.com/questions/15828329/how-can-i-get-sel-selector-from-object-file-mach-o-how-sel-stored-in-mac

\subsection{Disassembly}

In parallel to extracting the type hierarchy, we start disassembling the binary starting from each function boundary and then lift disassembled function bodies to a graph representation that can be further processed, as explained in the next section. Besides determining segment boundaries and reconstructing the actual instructions and basic blocks, this includes keeping track of cross-references, e.g. to resolve local variables and the use of constants.

\subsection{Call Graph Reconstruction}

\begin{figure}[tb]
\subfloat[CG created by IDA Pro\label{fig:cg-before}]{\includegraphics[width=.5\linewidth]{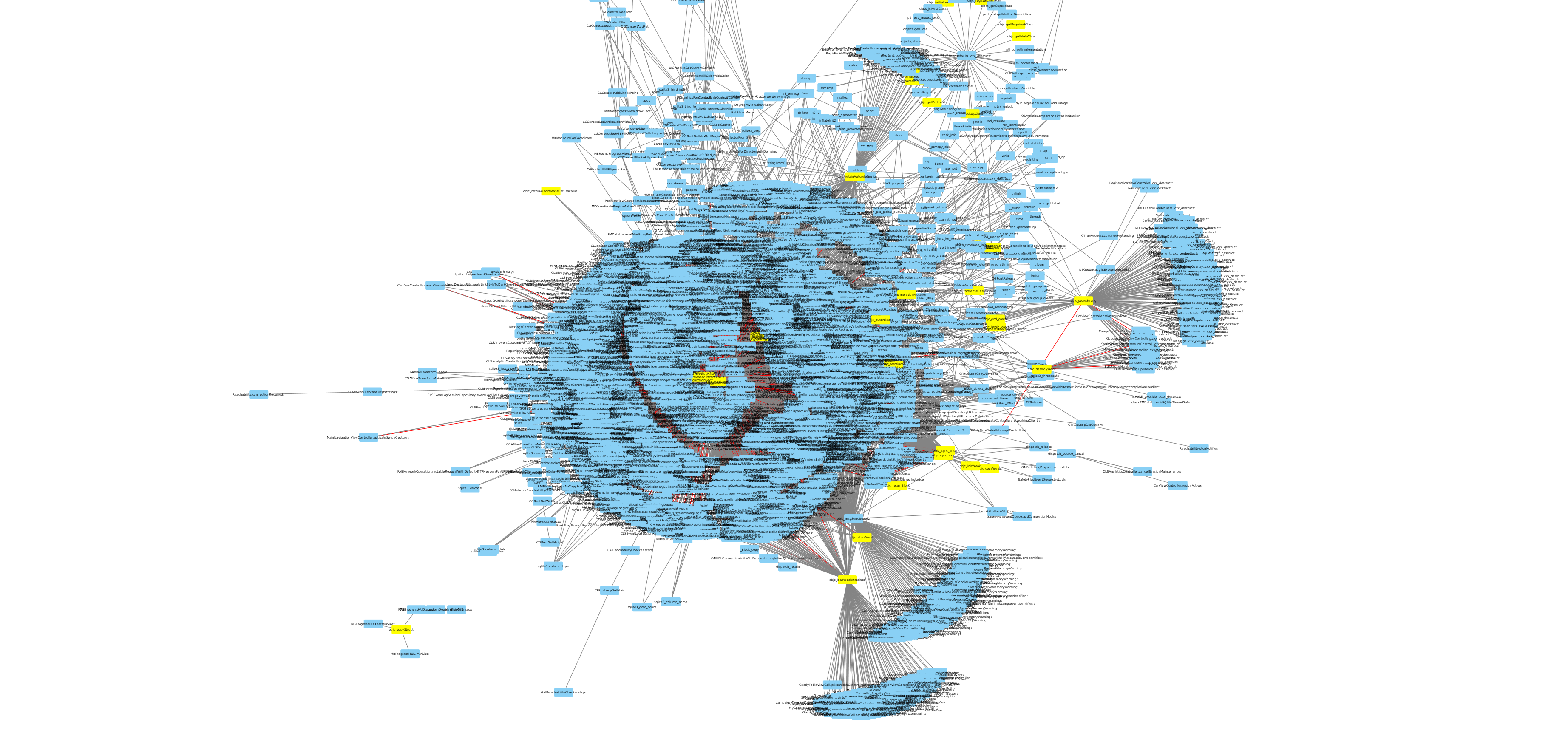}}\hspace*{-.1cm}
\subfloat[CG reconstructed by \thething{}\label{fig:cg-after}]{\includegraphics[width=.5\linewidth]{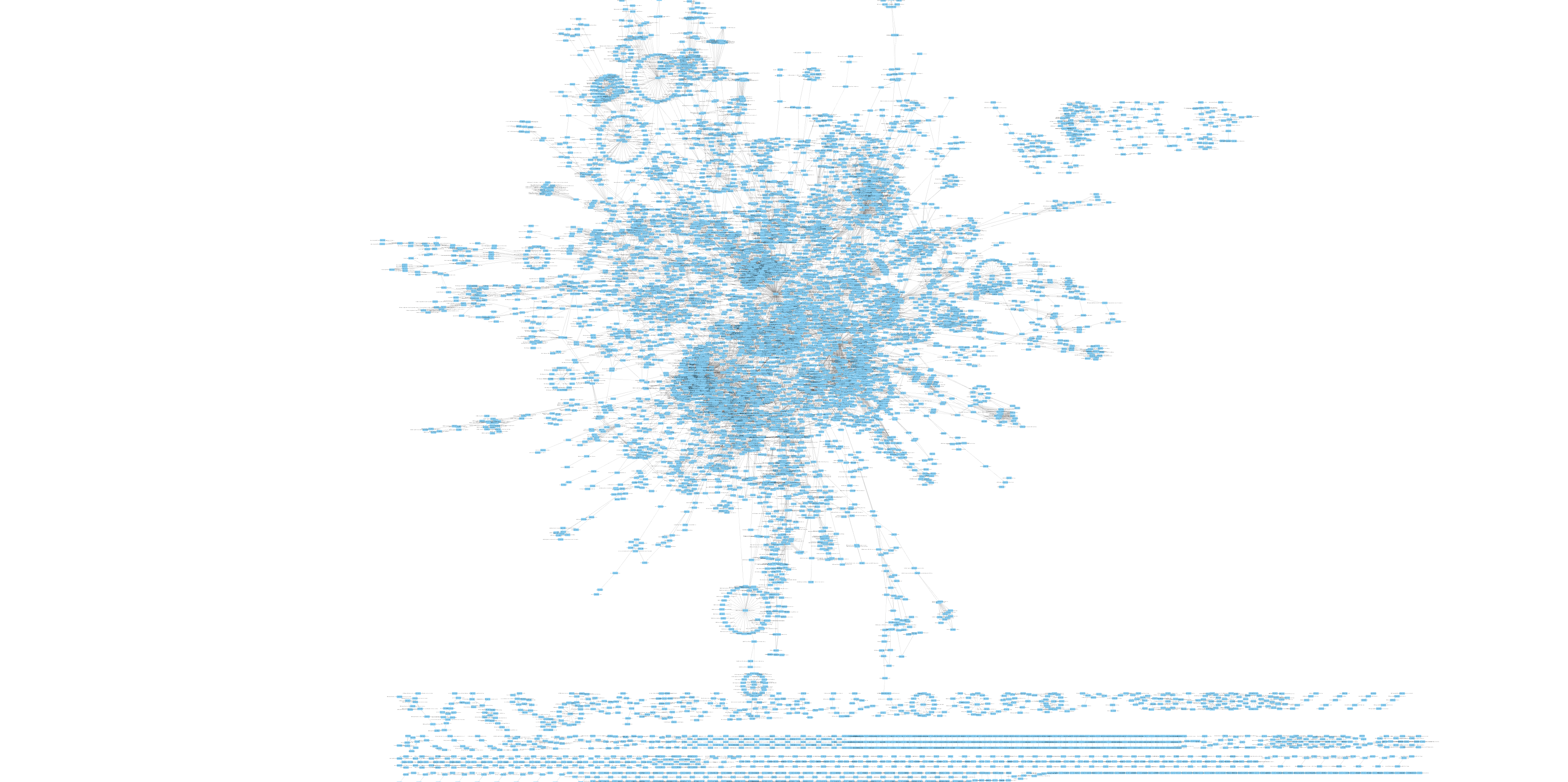}}
\caption{Call Graph before and after dispatch reconstruction\label{fig:cg}}
\end{figure}

Reconstructing a proper call graph is a prerequisite for any precise interprocedural analysis. Unfortunately, Objective-C has some quirks which make the reconstruction of a call graph not straightforward. As one of the first object oriented languages supporting dynamic binding and message-based dispatching, there is no such concept as a direct method call in Objective-C. Rather, the caller only constructs a message stating a receiver, a selector and optional arguments. This message is handed to the \texttt{objc\_msgSend} dispatcher of the Objective-C runtime which is responsible for finding the receiver, finding an appropriate method implementation matching the receiver, and performing the call. Thus, when naively creating a call graph from an Objective-C binary, one will receive a construct as shown in \autoref{fig:cg-before}. In fact, we are not aware of any disassembly tool (IDA Pro, Hopper, radare2) that is able to construct any other call graph representation than the one shown in \autoref{fig:cg-before}. A precise call graph creation requires to reconstruct the possible values of the receiver and selector argument to all \texttt{objc\_msgSend} calls. As these arguments can be dynamically set at runtime, possibly even by user-provided input values, generally reconstructing them by a static analysis is infeasible. In practice, however, we found that reconstruction of these arguments is possible with a high success rate, because the Objective-C/Swift compiler creates selectors as pointers to string constants and receivers as pointers to either the class or object instance. These are typically assigned to registers within the same function as the call to \texttt{objc\_msgSend}, however not necessarily near the call, nor always constructed in the same way. To reliably reconstruct these argument values, we thus employ the algorithm shown in Algorithm \ref{alg:methodcall} to reconstruct values of the registers \texttt{x0} (receiver) and \texttt{x1} (selector) for each call to \texttt{objc\_msgSend}.

\begin{algorithm}
\caption{Backtracing of registers for method call reconstruction\label{alg:methodcall}}
\begin{algorithmic}[1]
    
    \Procedure{Backtrace}{$reg$ , $\mathit{addr}$} \Comment{\emph{reg: register}}
    \State \Comment{\emph{addr: Code location with use of v}}
        \State$V \gets \{(reg,addr)\}$
        \While{$V \neq \emptyset$}
            \State$(x,l) \gets pop(V)$
            \State$instr \gets \textsc{InstructionAtLocation}(\mathit{l})$
            \If{instr is \textsc{Assignment}}
                \State$def \gets instr.\textsc{Definition}$
                \State$use \gets instr.\textsc{Use}$
                \If{$def$ matches $x$}
                    \State      \Comment{\emph{If immediate pointer to objc\_*-segments:}}
                    \If{\texttt{PointsToConstant}(use)}
                        \State{\textbf{return} $\{\textsc{Dereference}(use)\}$}
                    \Else
                        \State{$V \gets V \cup {(use,l)}$}
                    \EndIf
                \EndIf
            \ElsIf{instr is \textsc{BranchInstr}}
              \State$rcv \gets \textsc{backtrace}(x0, pred(\mathit{l}))$
              \State$sel \gets \textsc{backtrace}(x1, pred(\mathit{l}))$
              \State \textbf{return} \{rcv.sel\}
            \ElsIf{$pred(l) = \emptyset$}
            \State      \Comment{\emph{If reached begin of function:}}
              \If{$x = \texttt{x0}$}     
                \State \Comment{\emph{\texttt{x0} points to own class/instance name}}
                \State \textbf{return} \{\textsc{Self}\}
              \ElsIf{$x = \texttt{x1}$}  
                \State  \Comment{\emph{\texttt{x1} points to own function name (selector)}}
                \State \textbf{return} \{\textsc{SEL}\}
              \EndIf
            \Else
                \State$V\gets V \cup \{(x,l') | \mathit{l'} \in pred(\mathit{l}) \}$
            \EndIf
        \EndWhile   
    \EndProcedure
\end{algorithmic}
\end{algorithm}

The algorithm operates on a work list $V$ holding pairs of registers and code locations. It works backwards along the control flow graph and traces chains of register assignments until it reaches an assignment from a constant or the entry point of the function. The former case is handled by line 12, where registers hold a pointer to some struct in the Objective-C segments (such as \texttt{\_\_objc\_methname}). In the latter case, if register \texttt{x0} is not explicitly set at all (line 22), it refers to the first parameter of the function, which is typically a pointer to \textsc{Self}, i.e. the class or object declaring the function. Further, function calls can be constructed dynamically by retrieving receiver and selector as a return value from another function call (line 16). In this case, the algorithm will interprocedurally trace the values by recursively analyzing the called function. Although in theory soundness of the algorithm is limited in that it is only flow- but not path sensitive, our experience is that code created by the Objective-C compiler does not use path-dependent receiver and selector values.

Once receiver and selector have been reconstructed, the aforementioned class hierarchy is looked up to retrieve the address of the actual method implementation and an edge from the caller to that address is inserted into the call graph.

\section{A Supergraph Representation of Lifted Binaries}
\label{sec:supergraph}

The information gained during the reverse engineering process must be combined in a single representation serving the analyses. Traditional program analysis frameworks use fixed data structures with strong dependencies which cannot easily be extended and capture only a single level of abstraction (i.e., assembly, object-oriented constructs, or syntactical language-level issues). 
%This approach is straightforward and efficient, but it relies on all data structures to be successfully created before even starting the analysis. 
%That is, frameworks like Soot \cite{lam2011soot} and Frama-C \cite{cuoq2012frama} cannot operate on incomplete inputs, but require a complete and compiling code base to even start the analysis (work to infer missing information for source code-based analysis has been done in \cite{Dagenais2008,moonen2001generating}). 
For most use cases of static source code analysis, this is not a hurdle. For binary analysis, however, we must assume that the individual results of the reverse engineering steps are incomplete, inconsistent, or even missing. We thus choose to merge them in an extensible representation that allows to run analyses on incomplete information at different abstraction levels, obviously sacrificing completeness and -- depending on the analysis -- also soundness. While this might appear as a drawback to the reader, we point out that soundness and completeness are two extremes that cannot be achieved perfectly at the same time. For analyzing realistic programs, it is much more important to the user that the tool is able to operate on the program at all, while being ''as sound and complete as it gets'' (cf. \cite{listt2014,Machiry2017}).

\subsection{Construction of the Supergraph}

\thething{} represents the results of the reverse engineering step in form of a graph representation. The initial graph is constructed from the inputs of the reverse engineering steps (\emph{frontends}) and subsequently extended by further \emph{passes} to build a \emph{supergraph}. This supergraph is the basis for the actual analysis modules, which run \emph{graph traversal} queries against it. Graph models for static program analysis have been used before \cite{Yamaguchi2014,Martin2005,Janzen2003}, however only for approaches operating at source code level and with fixed graph structures. We apply this concept to binary analysis and allow the graph to be extended during analysis. We especially regard the latter aspect as crucial when lifting binaries from assembly to increasingly high-level representations, while coping with possibly incomplete results from each step.

%Yamaguchi et al. propose to represent a program in form of a Code Property Graph \cite{Yamaguchi2014} -- a combination of AST, Call Graph, Control Flow Graph and Program Dependence Graph. The benefit of this representation is that detecting vulnerabilities in an application is mere matter of defining a respective graph query. A similar graph-based approach has been proposed for the Program Query Language PQL \cite{Martin2005} and \cite{Janzen2003}.

% Explain the concept of a code property graph
A property graph $PG = (V, E, \Sigma, \sigma, \mathcal{L}, \ell)$ is a directed labeled ``multi-graph'' of vertices $V$ and edges $E = \{(v,u) \in V \times V\}$, where each vertex and edge has a (possibly empty) set of key-value \emph{properties}. $\Sigma$ denotes an alphabet of property keys for vertices and edges and the \emph{property function} $\sigma: (V \cup E) \times \Sigma \to S$ assigns values of value set $S$ to the property keys of vertices and edges. Further, each node and edge is assigned a \emph{label} by the labeling function $\ell: (V \cup E) \to \mathcal{L}$. The graph can be queried using \emph{graph traversal} functions. A graph traversal $T = (PG, \circ, \texttt{identity})$ is a monoid over a property graph $PG$ with an associative method chaining function $\circ$ and a neutral element \texttt{identity}. By chaining traversal functions, we can build arbitrary complex queries over nodes and edges and their labels and properties. We write $\circ^n$ to denote the $n$-times repetition of a traversal function and $\circ^*$ to denote its reflexive transitive closure. The extensible supergraph comprises different building blocks of an iOS app and as further frontend modules or passes are added to \thething{}, additional labeled nodes and edges may be added.
%The core concepts are defined by the following set $\mathcal{L}$ of labels and used as shown in \autoref{tab:props} and \autoref{tab:edges}.
% \begin{gather*}
%   \mathcal{L}  = \{   \textsc{Program}, \textsc{Function}, \textsc{Method}, \textsc{Class}, \\
%                       \textsc{BasicBlock}, \textsc{Instruction}, \textsc{Ivar} \}
% \end{gather*}

\begin{table}[tb]
  \caption{Node properties}
  \label{tab:props}
  \centering

  \begin{tabularx}{\linewidth}{l|l|X}
  \hline
  \textbf{Node Label ($\mathcal{L}$)} & \textbf{Property ($\Sigma$)} & \textbf{Description} \\
  \hline
  \textsc{Program}  & \texttt{ea} & Extended address \\
                    & \texttt{name} &  Name of binary\\
                    & \texttt{entltl} &  Entitlements of app\\
                    & \texttt{info} &  Content of Info.plist\\
  \hline
  \textsc{Function} & \texttt{ea} &  Extended address\\
                    & \texttt{name} &  Name of function\\
                    & \texttt{is\_ext} &  External or implemented function\\
                    & \texttt{llvm} &  LLVM IR representation of function body\\
                    & \texttt{is\_ep} &  Is function an entrypoint of the app?\\
  \hline
  \textsc{Method} & \texttt{name} &  Name of method\\
  \hline
  \textsc{Class}  & \texttt{name} &  Name of class\\
  \hline
  \textsc{BasicBlock} & \texttt{ea} &  Extended address\\
  \hline
  \textsc{Instruction}  & \texttt{ea} &  Extended address\\
                   & \texttt{bytes} & Bytes of the actual instruction\\
                   & \texttt{asm} & Assembly representation\\
  \hline
  \textsc{Ivar} & \texttt{ea} &  Extended address\\
  \hline
  \end{tabularx}
\end{table}

\begin{table}[tb]
  \caption{Edges}
  \label{tab:edges}
  \centering

  \begin{tabularx}{\linewidth}{p{4.5cm}|X}
  \hline
  \textbf{Edge} & \textbf{Description} \\

  \hline
  $\texttt{impl}\left(\textsc{Function} \times \textsc{Method}\right) $ & Method of a class that corresponds to a function\\
  $\texttt{succ}\left(\textsc{BasicBlock} \times \textsc{BasicBlock}\right) $ & Successor of basic block \\
  $\texttt{def}\left(\textsc{Instruction} \times \textsc{Instruction}\right) $ & Definition of a variable used by the current instruction \\
  $\texttt{implements}\left(\textsc{Function} \times \textsc{Method}\right) $ & Function implementing a method\\
  $\texttt{calls}\left(\textsc{Function} \times \textsc{Function}\right) $ & Function calls (possibly indirect) \\
  $\texttt{has\_superclass}\left(\textsc{Class} \times \textsc{Class}\right) $ & Class hierarchy \\
  $\texttt{has\_protocol}\left(\textsc{Class} \times \textsc{Protocol}\right) $ & Protocols of a class \\
  $\texttt{isa}\left(\textsc{Class} \times \textsc{Class}\right) $ & Metaclass hierarchy \\
    \texttt{has\_meth}$((\textsc{Class}$ $\cup$ $\textsc{Protocol}$) $\times\ \textsc{Method})$ & Method of a class/metaclass\\
  \hline
  \hline

  \hline
  \end{tabularx}
\end{table}

The root of the supergraph is a \textsc{Program} node, representing the Mach-O binary of the app. Its properties hold values such as entitlements and contents from the Info.plist file. \textsc{Program} has edges to all \textsc{Function}s implemented and imported by the binary. Imported external functions have a \textsc{is\_ext} property set and an address (\textsc{ea}) of -1, while functions implemented by the application have their function body stored in LLVM IR. While not every function corresponds to a method, the other way round applies: every \textsc{Method} of a \textsc{Class} that is implemented by the app corresponds to a respective function, denoted by the \texttt{implements} edge. Every function has edges to its basic blocks which have edges to their instructions. These instructions refer to the actual assembly instructions and thus map to specific addresses (\texttt{ea}), as opposed to the LLVM representation. We further capture data flows by adding use-def edges pointing from every use of a memory location or register to the instruction defining its value, whereas the edge is assigned a property \texttt{var} referring to the respective memory location.
% (which is not shown in \autoref{tab:edges}).
Consequently, this basic graph model captures the control flow graph (CFG), call graph (CG), and data dependence graph (DDG), together with additional information such as LLVM IR function bodies and application metadata. Just by processing the graph and without adding further information, this representation allows to construct a forward dominance tree (FDT) and a program dependence graph (PDG), which we do not explicitly compute and store for performance reasons. We will show in the example below in \autoref{sec:evaluation} how we can leverage this graph structure to compute static taint flows and data-dependent program slices, i.e. the minimal set of instructions in a function that has equivalent semantics with respect to the values of a specific register or memory location as the original function.

% Following the notation from An algebra and equivalences to transform graph patterns in Neo4

% $\mathcal{L}(v)$ is the label of node $v$.

% $p\_i(v)$ denotes property $i$ of node $v$.

% If $x$ is the attribute of a graph relation, $x.p$ denotes the property value.

% $E = \{\textsc{implementsMethod} \in \}$ \todo{?}

%TODO TIKZ picture of the graph's nodes and edges, colored by the "level"/"pass" they are created from

\subsection{Static Program Analysis using the Supergraph}

This section discusses the use of \emph{graph traversals} for building increasingly abstract static analyses in the form of queries against the supergraph. Just like passes can extract the graph to hold more high-level information, we start with basic queries which are then re-used in more complex patterns.

\subsubsection{Call Graph Traversals}
Functions serving as entry points to the application are marked by the boolean property \texttt{is\_ep}. 
%As stated in \autoref{ssec:entrypoint}, 
In contrast to traditional linux binaries, execution of an app does not start at a single main method, but at various callback handlers which must be considered at entry points into the call graph. A graph traversal querying for all vertices with the \texttt{is\_ep} flag set is rather simple:
\begin{align*}
\textsc{Entrypoints} = \{  v \in V :\ & \sigma(v, \texttt{is\_entrypoint} = \texttt{true})\}
\end{align*}
Further exploring the call graph, it will be interesting to find all functions that are called from a given function, i.e. all \emph{callees}. Considering that function calls in the graph representation already abstract away the specifics of the Objective-C runtime such as reflective invocations via \texttt{objc\_msgSend}, this simple graph traversal can already reveal interesting insights into an app, such as the iOS APIs used by it. We denote the query for all callees of a function $x$ as follows:
\begin{align*}
\textsc{Callees}(x) = \{  v \in V :\ & (x,v) \in E\ ,
\\ & \ell(v) = \textsc{Function}\ ,
\\ & \ell((x,v)) = \texttt{calls}\}
\end{align*}
By transitively chaining this graph traversal to get its transitive hull, it is now possible to create a more abstract graph traversal that gives all functions which are transitively reachable from a given function $x$:
\begin{align*}
\textsc{Reachables}\left(x\right) = \bigcup_{v \in V} x\ \circ^*  \textsc{Callees}(v)
\end{align*}
% Likewise, by negating the now available traversal \textsc{Reachables}, we can define a graph traversal returning all unreachable functions. This is for instance relevant to identify unused library code which is imported but never used by an application.
% \begin{align*}
% \textsc{Unreachables} = x \in V : x \not\in \textsc{Reachables}
% \end{align*}

\subsubsection{Control Flow Traversals}
Analog to call graph traversals, the control flow of each function implemented by the app can be explored. For instance, to retrieve all immediate successors of a basic block, i.e. typically the immediately following instruction and the jump target, a graph traversal \textsc{Successors} can be defined.
\begin{align*}
\textsc{Successors}\left(x\right) = \{v \in V : &    (x,v) \in\ E\ 
\\ & \text{and}\ \ell((x,v)) = \texttt{succ} \}
\end{align*}

Again, the query for an immediate successor can be transitively extended. Rather than querying for the mere set of all successors (i.e., all blocks of non-dead code in a function), we can retrieve all execution paths from a given basic block $x$. As functions with loops would result in infinite paths, a maximum path length $l_{max}$ can be set.
\begin{align*}
\textsc{ExePath}\left(x\right) =\ & \bigl<v_0, .., v_n : v_i, v_{i+1} \in V,
\\ & (v_i,v_{i+1}) \in E, v_0 = x\ ,
\\ & n < l_{max}\ ,
\\ & \ell((v_i,v_{i+1})) = \texttt{succ} \bigr>
\end{align*}

\subsubsection{Data Flow Traversals}
By following the \texttt{def} edges between instructions in a function, it is further possible to express a backward slicing of a function with respect to a value $q$ used at an instruction $x$:
\begin{align*}
\textsc{DataFlow}\left(x, q\right) =\ & \bigl<v_0, ..., v_n : v_i, v_{i+1} \in V, 
\\ & (v_i,v_{i+1}) \in E, v_0 = x\ , 
\\ & \sigma((v_0,v_1), \texttt{var}) = q\  ,
\\ & \ell((v_i,v_{i+1})) = \texttt{def} \bigr>
\end{align*}

\section{\thething{} Analysis Framework}\label{sec:framework}

In this section, we discuss the software components of \thething{} and point out how the aforementioned graph traversals can be made more accessible to the user in the form of a simple domain specific language (DSL). \thething{} uses a plugin-architecture to delegate the graph construction and analysis of \thething{} to several modules. It first creates an empty graph object and then calls a set of \emph{frontend} modules which operate on the binary file and create initial nodes and edges into the graph. \thething{} then calls a series of \emph{passes} -- modules which process the existing graph and create further nodes and edges, representing higher-level concepts. Besides extending the graph, passes implement static analyses in the form of graph traversals and optionally make them available as new graph traversals with a shorthand notation in the DSL. That is, a complex graph traversal for an intraprocedural static taint analysis, for instance, can be made available to the user as a simple \texttt{tainted(source,sink)} operation that returns all ``tainted'' paths from a specific source to a specific sink.
%Analysis modules can thus depend on each other and re-use graph traversals implemented by other modules, thereby allowing increasingly abstract analyses.
The query DSL is implemented on top of the Gremlin\footnote{\nolinkurl{https://tinkerpop.apache.org/gremlin.html}} graph query language.

\subsection{Frontends}

Frontends are responsible for extracting information from a binary and provide it in form of a property graph. \thething{} includes three main frontend modules: the lifter, a disassembly frontend, and a class hierarchy frontend.

\subsubsection{Lifter Frontend}

The lifter operates on the binary and reconstructs the program structure in form of a McSema CFG file\footnote{\nolinkurl{https://github.com/trailofbits/mcsema/blob/master/mcsema/CFG/CFG.proto}}. As McSema itself is not able to handle specifics of iOS apps, such as cross-references or the class hierarchy, we implemented a new lifter frontend that operates on the aarch64 Mach-O binary, creates the disassembly (using radare2), and reconstructs the control flow graph and cross-references to variables. The result is a McSema-compatible ``CFG'' file which is then loaded into a graph with nodes representing functions, basic blocks, instructions, variables, and cross-references between code and data locations.

\subsubsection{Class Hierarchy Frontend}

As the lifter is concerned with disassembly only, it does not have any understanding of the object-oriented concepts of Swift and Objective-C. While it merely operates at the level of functions and instructions, we obviously need further information about the class hierarchy which can be easily extracted from the segments of the Mach-O binary.
%, as explained above in \autoref{sec:backgroun}. 
However, none of the existing tools was able to provide a clean and complete representation of the Objective-C class hierarchy to our satisfaction. IDA Pro\footnote{version 6.8, in our case} does not have any understanding of object-oriented structures. \texttt{jTool} dumps only class names. radare2 creates a class hierarchy, but omits properties and protocols. \texttt{llvm-objdump} provides extensive output of the Objective-C segments, but omits the actual pointers to function entry points. We therefore implemented a module to parse the Objective-C sections from the Mach-O file and reconstruct classes, meta-classes, protocols, properties, and methods and add respective vertices and edges to the graph.

\subsubsection{Disassembly Frontend}

The lifter frontend already creates representations of functions, basic blocks and individual instructions. However, it does not parse instructions in any way and only represents them as \textsc{Instruction} nodes with edges to their basic block. The disassembly frontend adds semantics to instructions by parsing the disassembly using an ANTLR streaming parser \cite{parr1995antlr} and processing the disassembly as follows:

Whenever an instruction refers to a cross-reference, an \texttt{xref} edge to the respective target is inserted into the graph. In case of an immediate operand (i.e. a constant value) or when a direct memory reference to a constant could be reconstructed by radare2, the respective constant value is assigned as a property to the \texttt{xref}'ed node. This allows to directly find all uses of a specific constant in the supergraph and is a significant advantage when writing graph traversals.

When an instruction is a branch statement to the \texttt{objc\_msgSend} method, we reconstruct the values of the registers holding the \texttt{SEL} and \texttt{IMP} values using the algorithm from Algorithm~\ref{alg:methodcall} and insert a \texttt{call} edge into the graph. Direct branches to functions are obviously easier to handle and do not require any back tracing to create a respective \texttt{calls} edge. The output of this frontend is thus comprised of a call graph and cross-references that resembles the actual high-level program in Swift/Objective-C, rather than the immediate references at assembly level.

Further, the disassembly frontend creates use-def edges (named \texttt{def}) connecting the instruction using a memory location to the instructions defining it. As these edges are constructed at assembly level and not in a single static assignment form, a single \emph{use} instruction may point to several \emph{def} instructions in different basic blocks. As for memory locations, we support the 31 general purpose ARM registers, as well as references to memory locations, including simple pointer arithmetic on stack-relative offsets.

\subsection{Passes}

After creating the initial graph from the frontend's output, \emph{passes} are responsible for extending the graph by further edges, nodes, and properties. Although passes are thought of as a plugin-mechanism to add further graph transformations at a later time, \thething{} includes a built-in pass to link and extend the isolated sub-graphs from the three frontends. For instance, the lifter frontend detects function entries by looking up the values from the \texttt{LC\_FUNCTION\_STARTS} Mach-O header and gets its name by demangling the respective symbol. The class hierarchy frontend running in parallel takes a different approach and extracts method names from the respective binary segments. As these are more precise in that they do not require demangling and contain the actual method signature, we link functions to methods with a matching address by inserting \texttt{implements} edges between the respective nodes. This way, it becomes possible to query the graph for the precise method signature and retrieve the body of the function implementing the method -- either as McSema LLVM IR or as a subgraph of assembly instructions.

%\begin{figure}[tb]
    %\centering
    %\includegraphics[width=\columnwidth]{img/neo4j-easyjswebview.png}
    %\caption{Graph representation of methods (blue), basic blocks (purple), and string constants (yellow)}
    %\label{fig:neo4j-graph}
%\end{figure}

% \subsection{Graph Database}

% \begin{itemize}
%   \item Neo4J with Tinkerpop API. Could be deployed as a cluster
%   \item Some figures: Size of app, size of graph, time for disassembly, parsing, loading the graph, extending the graph.
% \end{itemize}

\subsection{Extensible Graph Query Language}

As a persistence layer, \thething{} uses the Neo4J graph database \cite{Webber:2012:PIN:2384716.2384777}. However, to abstract away the persistence layer, \thething{} does not directly access Neo4J, nor uses its built-in query language ``Cypher''. Rather, we use the Apache Tinkerpop framework that is a generic graph database interface and allows to easily exchange the backend e.g. by an in-memory database. Tinkerpop also comes with a database-independent graph traversal query language called \emph{Gremlin}. To make the supergraph more accessible for static analyses, \thething{} makes use of Gremlin's support for the creation of domain-specific languages (DSL). The purpose of the DSL is to wrap commonly used traversals in reusable shorthand notations that can be used to build increasingly complex queries. Consider for instance a graph traversal that returns all functions of an application which dynamically load and execute code. In plain gremlin, this query would be written as follows:

\begin{lstlisting}[language=Java,basicstyle=\footnotesize,breaklines=true,postbreak=\mbox{\textcolor{red}{$\hookrightarrow$}\space},numbers=left,
    stepnumber=1,label={lst:gremlin}]
g.V().hasLabel("FUNCTION")     // Get all functions
  .out("has_bb")               // and their basic blocks
  .out("instr")                // and their instructions
  .out("calls")                // which implement a call
  .has("name", "NSInvoke.invoke") // to [NSInvoke invoke]
\end{lstlisting}

In \thething{}'s domain specific language, the query can be written much simpler by calling the predefined shortcuts \texttt{functions} referring to all functions nodes and filtering them by the \texttt{calling()} shortcut:

\begin{lstlisting}[language=Java,basicstyle=\footnotesize,breaklines=true,postbreak=\mbox{\textcolor{red}{$\hookrightarrow$}\space},numbers=left,
    stepnumber=1,label={lst:dsl}]
functions().calling("NSInvoke.invoke")
\end{lstlisting}

To extend the DSL is such way, pass modules need to provide the implementation of a Gremlin \texttt{GraphTraversalSource} and a \texttt{GraphTraversal} object. The user can then write queries using the DSL and load it into the Jython-based interpreter of \thething{}.

%Evaluation:
\section{Spotting Vulnerabilities in iOS Apps}
\label{sec:evaluation}

%We download apps from https://iosninja.io/ipa-library
%We evaluated apps from \url{http://www.ipanati0n.com/}

% Lifting a (simple) example application and detecting a context-sensitive data flow in it

% Potential evaluations:
%
% 1)
% Apps which do not hide their preview thumbnail
% window.hide() on applicationWillResignActive and applicationDidBecomeActive
% 
% 2)
% Crypto
% -> has already been done in Automated Binary Analysis on iOS A Case Study on Cryptographic Misuse in iOS Applications
%
% 3)
% Code Execution via WebView delegate
% -> Haven't found an instance of this vulnerability in the wild
%
% 4)

In the following, we illustrate how \thething{} identifies vulnerabilities in iOS applications. For the sake of demonstration, we consider a remote code execution vulnerability in a JavaScript-Objective-C bridge -- a common technique mainly used by advertisement and analytics frameworks such as Google Analytics or MoPub, as well as web-based cross-platform development frameworks such as Apache Cordova or React Native.

\subsection{Example: Vulnerable WebView Delegates}

iOS apps use WebView components to render web contents such as HTML, JavaScript, and CSS from the Internet or from local resources in order to display advertisements or refer the user to online web pages without leaving the app. Cross-platform app development frameworks rely to a large extend on WebViews to render HTML-based user interfaces and use native APIs only to access lower-level functionality of the device. This is however not straightforward because web pages are confined in their WebView and cannot make direct calls into the iOS APIs. Although Apple does not provide an official way to provide iOS API access to JavaScript running in a WebView, there is a well-established workaround which is heavily used by applications throughout the app store (approx. 70\% of 4000 analyzed apps). It requires registering a delegate object of type \texttt{UIWebViewDelegate} at a WebView and implementing its method \texttt{shouldStartLoadWithRequest} which is invoked by the WebView whenever a new resource is about to be loaded from a URL.
Its successor \texttt{decidePolicyForNavigationAction} works likewise, and the vulnerability discussed herein is equally applicable.

The function receives the URL as an argument, can perform any action and finally return \texttt{YES} or \texttt{NO} to determine whether the WebView should load the new URL or discard the request. This can be used to call native functions from a WebView, by encoding the to-be-called functions in the URL, as illustrated by this line of JavaScript:
~\\

{\footnotesize \texttt{window.location="native:MYCLASS:do\_something:param1:param2";} }

~\\
By default, the WebView would try to load the URL \texttt{native:MY\-CLASS:do\_something:param1:param2} and obviously fail due to the unsupported URL scheme \texttt{native:}. To actually call native APIs, the app developer hooks into the WebView's lifecycle by implementing \texttt{shouldStartLoadWithRequest}, parsing the called URL, and mapping it to any native API calls. How this mapping is done remains up to the developer and is cause of a severe vulnerability if done incorrectly. If the implementation directly maps non-sanitized URL parameters to function calls, the app will call any native function as determined by the web page. In combination with either missing or flawed TLS communication or cross-site scripting vulnerabilities in the rendered web page, this opens a remote code execution vulnerability that allows an attacker to inject code into context of the running app and exfiltrate sensitive information from the device. Given the wide-spread use of this technique and the fact that apps can still opt out of transport security (ATS) (done in approx 75\% of 4000 analyzed apps), this is a realistic scenario.\\
Consider the implementation of \texttt{shouldStartLoadWithRequest} given in \autoref{lst:vuln}.

\lstset{language=[Objective]C}
\begin{lstlisting}[basicstyle=\footnotesize,caption=Vulnerable WebView delegate,breaklines=true,postbreak=\mbox{\textcolor{red}{$\hookrightarrow$}\space},numbers=left,xleftmargin=3em,stepnumber=1,framexleftmargin=10em,label={lst:vuln}]
- (BOOL)webView:(UIWebView *)webView shouldStartLoadWithRequest:(NSURLRequest *)request navigationType:(UIWebViewNavigationType)navigationType{
  
  NSString *req = [[request URL] absoluteString];
  if ([req hasPrefix:@"my-prefix:"]) {

   NSArray *split = [req splitSeparatedByString:@":"];    
   NSString* obj = (NSString*)[split objectAtIndex:1];
   NSString* method = [(NSString*)[split objectAtIndex:2]
              stringByReplacingPercent-EscapesUsingEncoding:NSUTF8StringEncoding];        
   SEL selector = NSSelectorFromString(method);
   Class cls = NSClassFromString(obj);
   NSMethodSignature* sig = [cls methodSignatureForSelector:selector];
   NSInvocation* invoker = [NSInvocation invocationWithMethodSignature:sig];
   invoker.selector = selector;
   invoker.target = cls;  
   [invoker invoke];    
   return NO;
  }
  
  if (! self.realDelegate){
    return YES;
  }
  
  return [self.realDelegate webView:webView shouldStartLoadWithRequest:request navigationType:navigationType];
}
\end{lstlisting}

This WebView handler parses the URL and extracts the strings \texttt{obj} and \texttt{method}, indicating the class and method to call. Without further sanity checks, a method call is constructed and then executed, imposing a way for attackers to trigger any code execution by manipulating the URL parameters of the WebView.

% Detecting the vulnerability
Detecting this vulnerability is not possible by a simple pattern matching, but rather requires analysis of intraprocedural control- and data flows, plus an investigation of the implemented protocols and the application transport security (ATS) settings of the app. More precisely, we must first check whether the app implements the \texttt{shouldStartLoadWithRequest} method of the \texttt{UIWebView} (and not only an equally named method). Within that method, we must trace data flows from the attacker-controlled input (here: the \texttt{request} argument) to sensitive functions (here: \texttt{NSClassFromString}, for instance) and then check whether the respective value is passed as an argument to the critical \texttt{[NSInvocation invoke]} method call -- two instances of a classic static taint analysis problem. 

\begin{figure*}[tb]
  \centering
  \includegraphics[width=.7\linewidth]{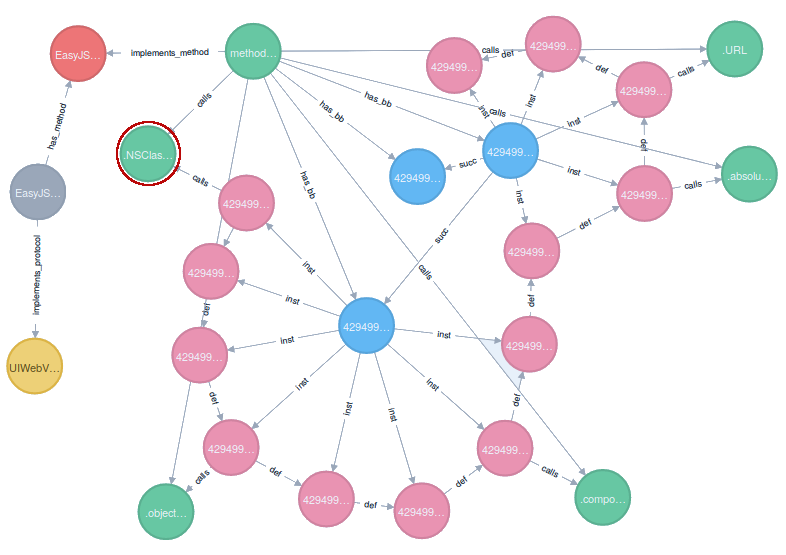}
  \caption{Supergraph of a vulnerable data flow, Instructions (pink), basic blocks (blue), functions (green), methods (red), classes (gray), and protocols (yellow)}
  \label{fig:neo4j}
\end{figure*}

%The idea of a taint analysis is to mark untrusted data -- either when it enters the program or as a result of a function call -- and to propagate the taint value until it reaches a critical sink. While the principle is simple, there are various challenges in designing a precise taint analysis that does not ''over-taint'' and is still able to reliably detect data flows. This is especially the case when analyzing binaries where the analysis has to keep track of pointer arithmetics and different types of memory locations and cannot rely on high-level constructs such as object instances, scopes, etc. Also, the definition of the taint propagation logic, i.e. the function that maps taint flags to the result of instructions operating on possibly tainted input values, is not fixed, but must rather be tailored to the specific type of analysis. 
To support such data flow analyses, we analyze \emph{use-def} chains in the supergraph. A use-def chain is a directed acyclic graph starting at an instruction that \emph{uses} a value to all instructions that previously \emph{define} the value. These chains are represented by \texttt{def} edges in the supergraph so that it is possible to trace back all data flows within a function, starting from a ''critical'' sink. This is illustrated by \autoref{fig:neo4j}. The figure shows a subgraph of the supergraph that captures the vulnerable pattern from \autoref{lst:vuln}, whereas the yellow node indicates the \texttt{UIWebView} protocol, the gray node the class implementing the protocol, the red node the method declared by that class, and the connected green node the function implementing that method. All further green nodes refer to \emph{external} functions, i.e. functions which are called, but not implemented by the binary. Blue nodes refer to the basic blocks of the function and pink nodes stand for individual instructions. The subgraph does not include all instructions of the function, but only the ones along the use-def-path from the critical data sink to the begin of the function -- i.e. to the point where they are populated from method arguments. This can be understood by starting at the green node with the red outline, which represents the call to \texttt{NSClassForName} (i.e., the sink) and then following the \texttt{def} arrows along all pink statement nodes to the head of the \texttt{shouldStartLoadWithRequest} method. This chain represents the unsanitized data flow from the method's argument to the critical \texttt{NSClassForName} call, spanning two basic blocks (created by the \texttt{if}-statement in line 4 of \autoref{lst:vuln}) and four calls to external functions before it enters the sink: \texttt{URL}, \texttt{absoluteString}, \texttt{componentsSeparatedByString}, and \texttt{objectAtIndex}. None of these functions is considered to sanitize the argument, so we end up with a critical function that directly operates on possibly malicious user input. The calls invoked by line 8 and 11 of the listing are not contained in the subgraph, because they are not part of the data flow relevant to reach \texttt{NSClassFromString} and thus not part of the program slicing.
Retrieving information about the ATS configuration of the application is trivial and can be done by evaluating the \texttt{info} property of the \textsc{Program} node. 
As the example shows, \thething{} is able to detect fairly complex vulnerability patterns that include data flow- and control flow-based program slicings, call graph patterns and configurations from metadata such as from \texttt{Info.plist}. This goes well beyond detecting simple calls to unsafe APIs or incorrectly ordered cryptographic operations.

\subsection{Performance and Footprint}

Similar to PQL \cite{Martin2005}, our approach separates the heavy lifting phase from the actual search for vulnerability patterns. This has the advantage that the computing-intense binary lifting and graph construction needs to run only once, while users may add further analyses at any time afterwards, and simply run them as queries against the already existing supergraph. However, we obviously had to achieve a trade-off between an extensive but slow pre-computation and a faster but limited supergraph creation. To give the reader an impression of performance and memory footprint of \thething{}, we consider the vulnerable Swift app from above.

The aarch64 Mach-O binary of the app itself is 301 KB large, implements six classes and a total of 85 functions -- a large portion of them being stubs and parts of the Objective-C runtime. Additionally, it makes use of 55 external functions which will result in further nodes and edges in the supergraph. 
Running \thething{} against the app results in a graph of 6800 nodes, 7340 edges and 13700 property values, totaling to a 5.4 MB graph database. The time to lift the binary amounts to 20 seconds, plus 24 seconds to construct the supergraph\footnote{on a standard x86 i7 laptop with 16 GB of RAM} and run the passes. It is reasonable to assume that this simple app denotes an upper bound for the ratio of graph database size to binary size and a lower bound for the total database size and computing time.

% ASsembly parsing + CG creation:   1371 ms
% Graph creation:                  24072 ms

%\todo{Vergleichen mit ''echter'' App, Ableiten von Aussage was die Grenzen sein werden}

\section{Related Work}
\label{sec:relawo}
Considerably less work on static analysis of iOS apps than on Android apps exists and as discussed in \cite{meng2016binary}, analyzing binaries imposes significantly higher challenges than source- or bytecode. Nevertheless, the problem has been tackled before by PiOS \cite{EgeleM.KruegelC.KirdaE.Vigna2011}, a privacy analysis approach for iOS binaries by Egele et al., which -- similar to our work -- strives for an automated reverse engineering and data flow analysis of iOS apps. Also, work by Feichtner et al. \cite{Feichtner2018} aims at identifying vulnerabilities in iOS binaries. Both differ in several aspects from \thething{}: PiOS hardcodes their data flow analysis directly on 32-bit ARM assembly and does thus not support newer platforms or an extension by vulnerability patterns. Feichtner et al. use dagger, a fork of the LLVM disassembler to lift ARM assembly to LLVM IR and immediately compute a forward program slicing with respect to the use of cryptographic functions. The result is a LLVM slice which is not executable and specific to the analyzed CCryptor API, but typically much smaller than the overall graph generated by \thething{}. In contrast to these works, \thething{} combines the result of different reverse engineering inputs, including disassembly and LLVM lifting, into a unified graph representation that tolerates missing information. For LLVM lifting, we extend McSema which in general aims at generating executable bitcode. Rather than detecting a specific vulnerability, \thething{} thus provides a generic iOS analysis framework that detects vulnerabilities specified by high-level queries.
Further, related work includes \emph{iRis}, an approach describing the reconstruction of Objective-C methods calls to detect the use of private APIs \cite{Deng2015} and \emph{Cricket} \cite{JakubBrecka2016}, a decompiler framework for Objective-C. The difficulty of analyzing a large corpus of realistic iOS apps has been addressed by CRiOS \cite{crios2016}, a framework for mass downloading and decrypting apps from the App Store, which achieved a success rate of \textapprox{}51\%.  % Mass download of >43k apps from iOS App Store. Only 50% of the apps can be dumped successfully, mainly consist of libraries, SSl servers are partly insecure
%From a practical perspective, dagger is currently limited to 32-bit platforms only, whereas Apple has made the transition from 32 to 64 architectures with the A7 SoC as of 2013.
 % \begin{itemize}
 %   \item They use dagger instead of McSema.
 %   \item No recovery of type hierarchy
 %   \item Not a generic framework, but rather a detection mechanism for broken cryptography
 % \end{itemize}
% weaknesses:
%   - points-to analysis is context-insensitive. As a result, polymorphism gives an incorrect call graph.
%   - is only aware of information from the binary, i.e. UI callbacks registered in .nib files (created by Interface Builder in XCode) are not considered
%Phasar\footnote{\url{https://phasar.org}} is related to our work in that it is an extensible static analysis framework operating on LLVM IR. It does however not consider binary lifting and will typically operate on LLVM IR created by the clang compiler. In that respect, Phasar is orthogonal to our approach and can be combined with our framework for more advanced static analysis of lifted binaries.

%\cite{Lerch2014}
%\cite{Mann2012} %Android

\section{Conclusion}
\label{sec:conclusion}

With \thething{}, we push forward the state of the art in static analysis frameworks for iOS apps. To overcome failing and incomplete lifting results, we do not rely on one specific intermediate representation but rather propose an extensible graph-based representation that is populated from various reverse engineering frontends. This representation tolerates missing information and allows static analyses in the form of graph traversal queries. In addition to the overall approach we made practical contributions such as the implementation of a radare2-based disassembly frontend for iOS binaries that fixes various shortcomings of existing lifters.
 By means of a simple data flow based vulnerability, we illustrated how \thething{} detects even complex vulnerability patterns using graph traversals. Part of our future work will be to increase soundness of our analysis by investigating the modeling of context-sensitive flows and moving away from generic graph databases to a performance-optimized graph model.

\bibliographystyle{splncs04}
\bibliography{references}

\begin{thebibliography}{10}
\providecommand{\url}[1]{\texttt{#1}}
\providecommand{\urlprefix}{URL }
\providecommand{\doi}[1]{https://doi.org/#1}

\bibitem{Bao2014}
Bao, T., Burket, J., Woo, M., Turner, R., Brumley, D.: Byteweight: Learning to
  recognize functions in binary code. In: Proceedings of the 23rd USENIX
  Conference on Security Symposium. pp. 845--860. SEC'14, USENIX Association,
  Berkeley, CA, USA (2014),
  \url{http://dl.acm.org/citation.cfm?id=2671225.2671279}

\bibitem{Deng2015}
Deng, Z., Saltaformaggio, B., Zhang, X., Xu, D.: {iRiS: Vetting Private API
  Abuse in iOS Applications}. Proceedings of the 22nd ACM SIGSAC Conference on
  Computer and Communications Security (CCS) pp. 44--56 (2015).
  \doi{10.1145/2810103.2813675},
  \url{http://dl.acm.org/citation.cfm?doid=2810103.2813675}

\bibitem{Deshotels2016}
Deshotels, L., Deaconescu, R., Chiroiu, M., Davi, L., Enck, W., Sadeghi, A.R.:
  {SandScout}. In: Proceedings of the 2016 ACM SIGSAC Conference on Computer
  and Communications Security - CCS'16. pp. 704--716. ACM Press, New York, New
  York, USA (2016). \doi{10.1145/2976749.2978336},
  \url{http://dl.acm.org/citation.cfm?doid=2976749.2978336}

\bibitem{dwarf2017}
{DWARF Debugging Information Format Committee}: {DWARF Debugging Information
  Format Specification Vesion 5.0} (2017)

\bibitem{EgeleM.KruegelC.KirdaE.Vigna2011}
{Egele, M., Kruegel, C., Kirda, E., Vigna}, G., Egele, M., Kruegel, C., Kirda,
  E., Vigna, G.: {PiOS: Detecting Privacy Leaks in iOS Applications.} NDSS
  p.~11 (2011)

\bibitem{Feichtner2018}
Feichtner, J., Missmann, D., Spreitzer, R.: {Automated Binary Analysis on iOS}.
  In: Proceedings of the 11th ACM Conference on Security {\&} Privacy in
  Wireless and Mobile Networks - WiSec '18. pp. 236--247. ACM Press, New York,
  New York, USA (2018). \doi{10.1145/3212480.3212487},
  \url{http://dl.acm.org/citation.cfm?doid=3212480.3212487}

\bibitem{Flirt}
Hexrays: {IDA F.L.I.R.T. Technology: In-Depth} (2015),
  \url{https://www.hex-rays.com/products/ida/tech/flirt/in_depth.shtml}

\bibitem{JakubBrecka2016}
{Jakub B\v{r}e{\v{c}}ka}: {A decompiler for Objective-C}. Ph.D. thesis,
  Univerzita Karlova (2016),
  \url{https://is.cuni.cz/webapps/zzp/detail/143329/?lang=en}

\bibitem{Janzen2003}
Janzen, D., {De Volder}, K.: {Navigating and querying code without getting
  lost}. In: Proceedings of the 2nd international conference on Aspect-oriented
  software development - AOSD '03. pp. 178--187. ACM Press, New York, New York,
  USA (2003). \doi{10.1145/643603.643622},
  \url{http://portal.acm.org/citation.cfm?doid=643603.643622}

\bibitem{Machiry2017}
Machiry, A., Spensky, C., Corina, J., Stephens, N., Kruegel, C., Vigna, G.:
  {DR. CHECKER: A soundy analysis for Linux kernel drivers}. USENIX Security
  Symposium pp. 1007--1024 (2017),
  \url{https://www.usenix.org/conference/usenixsecurity17/technical-sessions/presentation/machiry}

\bibitem{Martin2005}
Martin, M., Livshits, B., Lam, M.S., Martin, M., Livshits, B., Lam, M.S.:
  {Finding application errors and security flaws using PQL}. ACM SIGPLAN
  Notices  \textbf{40}(10), ~365 (oct 2005). \doi{10.1145/1103845.1094840},
  \url{http://portal.acm.org/citation.cfm?doid=1103845.1094840}

\bibitem{meng2016binary}
Meng, X., Miller, B.P.: Binary code is not easy. In: Proceedings of the 25th
  International Symposium on Software Testing and Analysis. pp. 24--35. ACM
  (2016)

\bibitem{crios2016}
Orikogbo, D., B\"{u}chler, M., Egele, M.: Crios: Toward large-scale ios
  application analysis. In: Proceedings of the 6th Workshop on Security and
  Privacy in Smartphones and Mobile Devices. pp. 33--42. SPSM '16, ACM, New
  York, NY, USA (2016). \doi{10.1145/2994459.2994473}

\bibitem{parr1995antlr}
Parr, T.J., Quong, R.W.: Antlr: A predicated-ll (k) parser generator. Software:
  Practice and Experience  \textbf{25}(7),  789--810 (1995).
  \doi{10.1002/spe.4380250705}

\bibitem{listt2014}
Rawat, S., Mounier, L., Potet, M.: {LiSTT: an investigation into
  unsound-incomplete yet practical result yielding static taintflow analysis}.
  In: Software Assurance Workhop (SAW 2014) (2014)

\bibitem{Wang2017}
Wang, S., Wang, P., Wu, D.: {Semantics-aware machine learning for function
  recognition in binary code}. In: Proceedings - 2017 IEEE International
  Conference on Software Maintenance and Evolution, ICSME 2017 (2017).
  \doi{10.1109/ICSME.2017.59}

\bibitem{wang2013jekyll}
Wang, T., Lu, K., Lu, L., Chung, S.P., Lee, W.: Jekyll on ios: When benign apps
  become evil. In: USENIX Security Symposium. vol.~78 (2013)

\bibitem{Webber:2012:PIN:2384716.2384777}
Webber, J.: A programmatic introduction to neo4j. In: Proceedings of the 3rd
  Annual Conference on Systems, Programming, and Applications: Software for
  Humanity. pp. 217--218. SPLASH '12, ACM, New York, NY, USA (2012).
  \doi{10.1145/2384716.2384777},
  \url{http://doi.acm.org/10.1145/2384716.2384777}

\bibitem{Yamaguchi2014}
Yamaguchi, F., Golde, N., Arp, D., Rieck, K.: In: 2014 IEEE Symposium on
  Security and Privacy. \doi{10.1109/SP.2014.44}

\end{thebibliography}

%\section{Vulnerable Objective-C/JavaScript bridge}

\end{document}